\documentclass[%
 reprint,
 amsmath,amssymb,superscriptaddress,
longbibliography,
prl
]{revtex4-1}
\usepackage{xparse}

\usepackage{appendix}
\usepackage{graphicx}
\usepackage{dcolumn}
\usepackage{bm}
\usepackage{hyperref}

\renewcommand{\d}[2]{\ensuremath{\frac{\text{d} #1}{\text{d} #2}}}

\newcommand{\ket}[1]{\ensuremath{\left| #1 \right>}}

\renewcommand{\vec}[1]{\mathbf{#1}}

\newcommand{\beginsupplement}{
\clearpage
\pagebreak
\setcounter{equation}{0}
\setcounter{figure}{0}
\setcounter{table}{0}
\setcounter{page}{1}
\makeatletter
\renewcommand{\theequation}{S\arabic{equation}}
\renewcommand{\thefigure}{S\arabic{figure}}
\onecolumngrid
\section*{\large{Supplemental Material}}
}
\usepackage{bbold}

\begin{document}
\title{Photon-mediated Peierls Transition of a 1D Gas in a Multimode Optical Cavity}

\author{Colin Rylands}
\affiliation{Joint Quantum Institute and
 Condensed Matter Theory Center, University of Maryland, College Park, MD 20742, USA}
 	\author{Yudan Guo}
\affiliation{Department of Physics, Stanford University, Stanford, CA 94305, USA}
\affiliation{
E. L. Ginzton Laboratory, Stanford University, Stanford, CA 94305, USA}
	\author{Benjamin L.  Lev}
\affiliation{Department of Physics, Stanford University, Stanford, CA 94305, USA}
\affiliation{
E. L. Ginzton Laboratory, Stanford University, Stanford, CA 94305, USA}
\affiliation{Department of Applied Physics, Stanford University, Stanford, CA 94305, USA}
\author{Jonathan Keeling}
\affiliation{SUPA, School of Physics and Astronomy, University of St Andrews, St Andrews KY16 9SS UK}
\author{Victor Galitski}
\affiliation{Joint Quantum Institute and
	Condensed Matter Theory Center, University of Maryland, College Park, MD 20742, USA}
\date{
    \today
}

\begin{abstract}
The Peierls instability toward a charge density wave is a canonical example of phonon-driven strongly correlated physics and is intimately related to topological quantum matter and exotic superconductivity.  We propose a method to realize an analogous photon-mediated Peierls transition, using a system of one-dimensional tubes of interacting Bose or Fermi atoms trapped inside a multimode confocal cavity.  Pumping the cavity transversely engineers a cavity-mediated metal--to--insulator transition in the atomic system. For  strongly interacting bosons in the Tonks-Girardeau limit, this transition can be understood (through fermionization) as being the Peierls instability. We extend the calculation to finite values of the interaction strength and derive analytic expressions for both the cavity field and mass gap. They display nontrivial power law dependence on the dimensionless matter-light coupling.
\end{abstract}
\date{\today}

\maketitle


\textit{Introduction} ---
The interaction between electrons and phonons has traditionally played a leading role in the formation of quantum phases of matter, with superconductivity being a prime example. Quantum simulation in optical lattices provides an enticing platform to explore new phases~\cite{Bloch}, but phonon-driven physics lies beyond traditional optical lattice capabilities as they are externally imposed and rigid.  The use of high-finesse optical cavities has been suggested as a route to overcome this by making the optical lattice fully dynamical and compliant~\cite{Lewenstein2006:Travelling,Gopalakrishnan1,Gopalakrishnan2}.  This requires cavities that support multiple degenerate modes, as single-mode cavities only allow dynamics of the lattice intensity, not its period. That is, while single-mode cavities have provided access to a diverse array of exotic quantum phenomena including self-organization~\cite{DomokosRitsch}, supersolids~\cite{Leonard}, spinor density-wave polariton condensates~\cite{Kroeze:2018fp}, dynamical Mott insulators~\cite{Klinder:2015jk,Landig:2016il}, and dynamical spin-orbit coupling~\cite{Kroeze:2019ex}, only multimode cavities support fully emergent optical lattices whose amplitude \textit{and} periodicity may vary~\cite{Lewenstein2006:Travelling,Gopalakrishnan1,Gopalakrishnan2}.  Multimode cavity experiments have already engineered a variety of photon-mediated interatomic interactions~\cite{Kollar, Vaidya, GuoVaidya, GuoKroeze}.  These could  lead to the creation  of new many-body systems and states of matter such as quantum liquid crystals made of photons and superfluid atoms~\cite{Gopalakrishnan1,Gopalakrishnan2} and superfluids exhibiting Meissner-like effects~\cite{Ballantine:2017dr}.

As we will show below, confocal multimode cavities coupled to one dimensional (1D) quantum gases provide a way to realize controllable electron-phonon-like interactions using ultracold atoms. Other proposals to study this physics include coupling fermionic atoms to an optical waveguide~\cite{fraser2019topological}, or to a crystal of trapped ions~\cite{Bissbort2013:Emulating}. One-dimensional ultracold gases also allow one to explore pairing physics with bosons, as resonant atomic collisions provide a knob to make bosons strongly repel~\cite{Chin:2010kl,Haller:2010dj}, even to the point that they behave like fermions~\cite{Paredes,Kinoshita,Haller:2009jrb}.  For such systems, the addition of \textit{attractive} interactions  can cause dramatic effects. Indeed, for 1D systems, even \textit{weakly} attractive interactions result in instabilities leading to strong correlations such that  the quasiparticle picture breaks down and collective modes emerge~\cite{Giamarchi, GogolinNerseyanTsvelik,Tsvelik}. A paradigmatic example is the Peierls instability that occurs because the susceptibility of a free Fermi gas diverges due to infinitesimal density perturbations at twice the Fermi wavevector~\cite{peierls}. If free phonons exist, then it is possible for the system to create an emergent lattice that matches this wavevector. Due to the diverging susceptibility, the system undergoes a metal-insulator transition and dynamically generates a mass gap. The Peierls transition is a canonical example of phonon-driven physics and intimately related to the continuum Su-Schrieffer-Heeger (SSH) model of 1D topological insulators in 1D~\cite{SuShreiferHeeger}.  Numerical work has shown the analogue of a Peierls transition for atoms in an optical lattice, where intersite hopping of bosonic atoms is modulated by the spin state of a second species of atoms~\cite{GonzalevCuadra2018:Strongly}.

In this Letter, we  show that a Peierls instability occurs in a strong, repulsively interacting 1D bosonic gas trapped inside a transversely pumped confocal multimode optical cavity. Building on demonstrated experimental capabilities~\cite{Vaidya,GuoVaidya}, we predict that by tuning the interatomic interactions to the hard-core, Tonks-Girardeau (TG) limit~\cite{Tonks, Girardeau}, the cavity can mediate a Peierls transition in the bosonic gas, with a mass gap and photon amplitude that is exponential in the matter-light coupling.  By using bosonization, we then extend these calculations to finite values of the interatomic interaction, as well as to interacting fermionic systems. In agreement with Ref.~\cite{GonzalevCuadra2018:Strongly}, we show that in these cases, the cavity can mediate a metal-insulator transition, albeit one of different character.  Moreover, we show that the dynamically generated mass gap and photon amplitude have a nontrivial power law dependence on the matter-light coupling. Self-organization of fermions in a single-mode cavity has also been previously discussed theoretically~\cite{KeelingBhaseenSimons,PiazzaStrack,ChenZhenuaZhai}, but here the diverging susceptibility requires the single cavity mode and Fermi wavevectors to match. Experimentally realizing a \textit{photon-mediated} Peierls transition would open new avenues toward exploring the role of Fermi surface nesting and charge density waves in exotic superconductors in simulators operating in a quantum-optical, many-body context.


\begin{figure}[t!]
	\includegraphics[width=\linewidth]{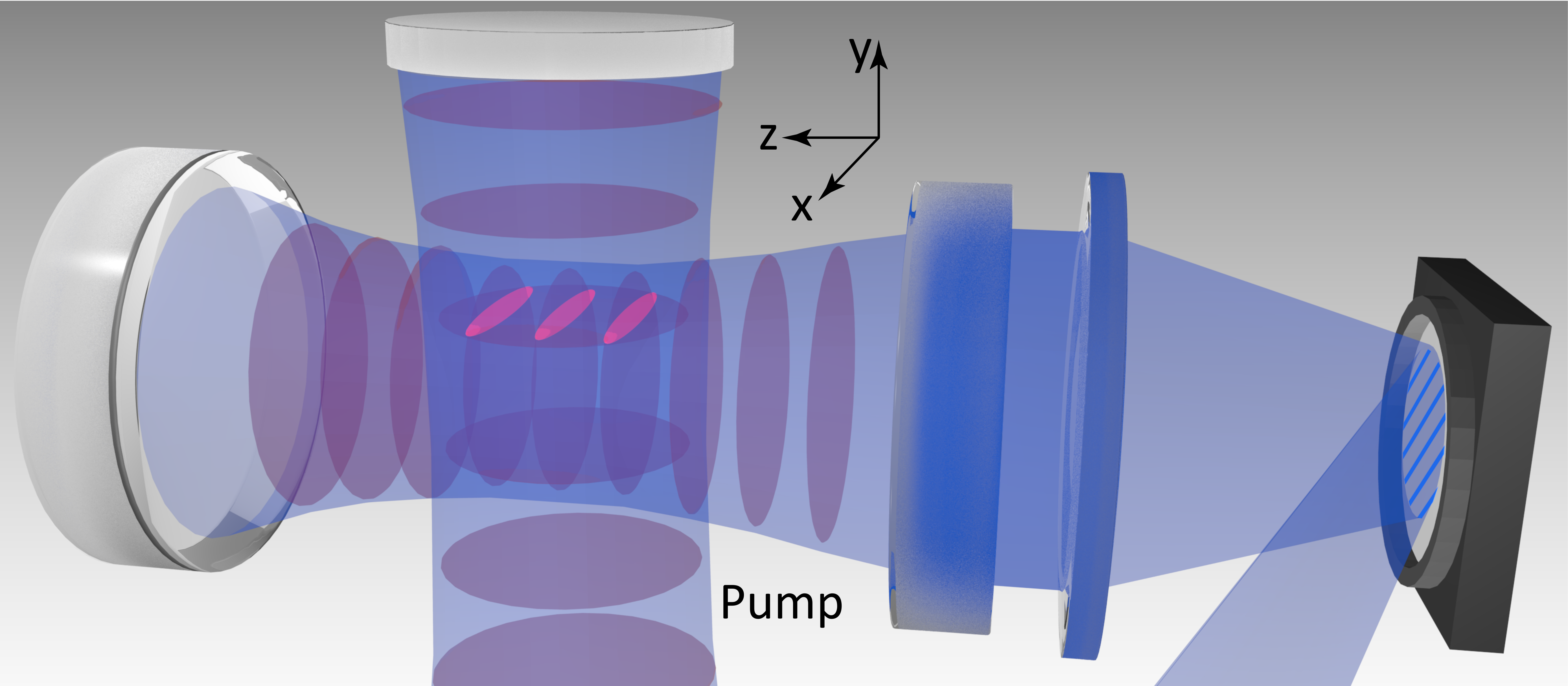}
	\caption{Schematic of our system. A gas of interacting  atoms is placed in a confocal optical cavity that supports many degenerate spatial modes. A transverse pump (blue) along $\hat{y}$ scatters photons off the atoms (red) into the the multimode cavity field (blue). The gas is confined by a 2D optical lattice (purple) and offset  from $y=0$ to avoid mirror-image interactions~\cite{Vaidya}. The amplitude and phase of the light is imaged via holographic reconstruction of a spatial heterodyne signal formed by interfering part of the pump with the emission. } \label{Schematic}
\end{figure}

\textit{Model} ---
The system considered,  depicted in Fig.~\ref{Schematic}, consists of 1D tubes of atoms placed in a transversely pumped confocal optical cavity.  As already noted, multiple optical modes are needed to allow a fully emergent optical lattice.  For true multimode operation, these modes must be degenerate or near degenerate.  A confocal cavity is the simplest stable resonator  allowing such degeneracy~\cite{siegman1986lasers}. 
To achieve a 1D trap geometry and uniform atom-cavity coupling, we confine  bosonic atoms in a strong $\lambda_T$-periodic 2D optical lattice formed by a retroreflected beam along the $\hat{y}$ pump direction and an intracavity standing wave along the $\hat{z}$ cavity axis.    A pump field along  $\hat{y}$ has a wavevector $k_r$ close to a  cavity resonance. By choosing the tube--lattice period such that $k_r \lambda_T/2\pi$ is an integer, the tubes lie at the peaks of the pump and cavity standing-wave fields so the atoms coherently Bragg scatter light into the cavity.  As a result, in contrast to experiments on self-organization~\cite{BaumannGuerlinBrenneckeEsslinger}, there is no spontaneous atomic organization in the $yz$ plane; rather, the atoms superradiantly emit into the cavity regardless of pump strength.  We choose the tubes to be near the cavity midplane $z=0$, and the long Rayleigh range of a confocal cavity ensures that the tubes at different $z$ will behave identically.  In $\hat{y}$, we choose all tubes to be centered at the same $y$ since tubes at different $y$ decouple, as discussed in the Supplemental Material~\cite{Supplement}. As such, we will describe the atoms in  tube $t$ through a bosonic field $\Psi_t(x)$, varying only along the free $x$ direction.
Degenerate confocal cavities with BECs in optical traps are practicable with existing technology~\cite{GuoKroeze}.

We can  write the Hamiltonian of the system as follows~\cite{Supplement}:
\begin{multline}
 \label{H}
 H=H_\text{cav}+ \int \mathrm{d}x \sum_{t=1}^{N_z} \Bigg\{ \Psi_t^\dag(x)\left[-\frac{\hbar^2}{2m}\partial_x^2-\mu\right]\Psi_t(x)\\
+ U \Psi_t^\dag(x)\Psi_t(x)\Psi_t^\dag(x)\Psi_t(x)-g \, \Phi(x)\rho_t(x)\Bigg\}.~~
\end{multline}
The first term, $H_\text{cav}=\hbar\sum_{\alpha,\nu}\omega_{\alpha,\nu} a^\dag_{\alpha,\nu}a_{\alpha,\nu}$, describes the cavity photons, where $\omega_{\alpha,\nu}$ is measured with respect to the transverse pump frequency. We sum  only over cavity modes that are  near resonant with the pump.  Modes are labeled by the longitudinal index $\alpha$ and transverse index  $\nu$. The second and third terms describe  atoms with mass $m$, chemical potential $\mu$, and a contact interaction of strength $U$.  We sum over an array of $N_z$ tubes, labeled by $t$, positioned in an array along   $\hat{z}$. The last term describes the coupling of the atomic density $\rho_t(x)=\Psi_t^\dag(x)\Psi_t(x)$ to the cavity photons, as induced by the pump. This term describes how the atomic density scatters photons between the transverse pump and the cavity modes. As described in Ref.~\cite{Supplement}, this term can be derived by adiabatically eliminating excited states of the atoms, yielding an effective AC light shift.  The photon field  is written as a sum over cavity modes
$\Phi(x)=\sum_{\alpha,\nu} c_\nu^\alpha\,\tilde{\Xi}_\nu(x) (a_{\alpha,\nu}^\dag+a_{\alpha,\nu}^{})$. The transverse mode functions  $\tilde{\Xi}_\nu(x)$  are found by taking the eigenmodes of the cavity---Gauss--Hermite  functions of order $(l_\nu,m_\nu)$ in the $x$ and $y$ directions, respectively---and convolving these with the Gaussian tube profile in the $y$ direction; see Ref.~\cite{Supplement}. The factors $c_\nu^\alpha$ come from the longitudinal spatial mode profile evaluated at the atom positions, and are discussed below. 
As noted above, this final term is a source for cavity photons independent of the density profile $\rho_t(x)$---the $\lambda-T$ tube spacing causes atoms to coherently scatter the pump into the cavity with intensity $\propto N^2_z$~\cite{kirton2019introduction}. The prefactor $g=\hbar g_0 \Omega/\Delta_a$ is the effective matter-light coupling where $g_0$ is the bare coupling, $\Omega$ the pump Rabi frequency, and  $\Delta_a$ is the pump-atom detuning.

As the Gauss--Hermite functions form a complete basis set, one might expect that $\Phi(x)$ could take any spatial profile. This would allow the cavity light  to match the atomic density, inducing a local interaction.  There are complications, however. First, while transverse modes become degenerate at confocal resonances of the cavity, they do so in alternating odd and even families, set by the parity of $n_\mu = l_\mu + m_\mu$.  We assume even $n_\mu$ hereon. Secondly, the factors $c_\nu^\alpha$  modify the sum over modes. As shown in Ref.~\cite{GuoVaidya}, the longitudinal mode profile assumes a form $c_\nu^\alpha=\cos{(\xi^\alpha-\pi n_\nu/4)}$, where $\xi^\alpha$ depends on which family we consider. This includes the effects of the Gouy phase~\cite{siegman1986lasers}, leading to an $n_\nu$ dependence.  If we consider the special case where $\xi^\alpha$ is a multiple of $2\pi$, we then see that for successive, even $n_\mu$, the factor $c_\nu^\alpha$ is sequentially $1,0,-1,0$.  The missing Gauss--Hermite functions prevent $\Phi(x)$ from obtaining an arbitrary form; equivalently, this yields a nonlocal photon-mediated interaction.  This can be nulled by using two pumps resonant with families  $\xi^\alpha=0$ and $\pi/2$, yielding a local interaction~\cite{GuoVaidya}.

Both the matter-light coupling $g$ and interatomic interaction strength $U$ are experimentally tunable parameters~\cite{Bloch}. In particular, the system may be tuned into the TG regime, i.e., $\gamma \equiv mU/\hbar^2\rho_0\to\infty$, using tight trapping and collisional resonances, where $\rho_0$ is the average 1D density~\cite{Kinoshita, Paredes}. The atoms behave like free fermions in this limit~\cite{Tonks,Girardeau} and so will exhibit a Peierls instability. Even strongly repulsive bosons away from the TG limit exhibit this instability.

\textit{Steady state} ---
We investigate the Peierls instability using a mean-field description of the photon field. We therefore consider the  equations of motion for the expectation of photon operators:
\begin{equation}
 \left<\dot{a}^\dag_{\alpha,\nu}\right>=(i\omega_{\alpha,\nu} -\kappa)\left<a^\dag_{\alpha,\nu}\right>
 -i\frac{g}{\hbar} \sum_t \int \mathrm{d}x c_\nu^\alpha\tilde{\Xi}_\nu(x)\left<\rho_t(x)\right>,
\end{equation}
where we have included a term $\propto\kappa$ accounting for cavity losses. Assuming a steady state, the mean field is
\begin{equation}\label{Phi}
 \left<\Phi(x)\right>=\int\!\mathrm{d}x^\prime\!\! \sum_{\alpha,\nu,t}\frac{2g  \omega_{\alpha,\nu}(c_\nu^\alpha)^2}{\hbar(\omega_{\alpha,\nu}^2+\kappa^2)}
 \tilde{\Xi}_\nu(x) \tilde{\Xi}_{\nu}(x^\prime)
 \left<\rho_{t}(x^\prime)\right>.
\end{equation}
For simplicity, we consider  the perfectly degenerate  limit $\omega_{\alpha,\nu}=\omega$.  Pumping two families as discussed above, $\sum_\alpha (c_\nu^\alpha)^2 = 1$, allowing for the explicit evaluation of the sum over modes: $\sum_{\nu}\tilde{\Xi}_\nu(x)\tilde{\Xi}_\nu(x^\prime)=\frac{w_0^2}{2 \sqrt{4 \pi}  \sigma_T}\delta(x-x^\prime)$\footnote{See Ref.~\cite{GuoVaidya} for near-degenerate case where the interaction becomes finite in range; Peierls physics remains the same as long as the interaction range is less than the intertube spacing.}.  Here, $\sigma_T$ is the transverse width of an individual tube, $w_0$ is the beam waist, and we made a simplifying assumption that the tubes are in the upper half of the cavity, $y>0$, to avoid mirror-image interactions~\cite{Vaidya}.  As all tubes behave identically, we can insert this into Eq.~\eqref{Phi} to give:
\begin{equation}\label{Phi2}
g \left<\Phi(x)\right>=\pi\hbar v_F \eta \left<\rho(x)\right>,
\quad
\eta\equiv\frac{g^2 N_z \omega w^2_0}{\sqrt{4 \pi} \sigma_T \pi \hbar^2v_F(\omega^2+\kappa^2)},
\end{equation}
where we have defined the dimensionless matter-light coupling $\eta$ and $v_F=\pi\hbar\rho_0/m$ is the Fermi velocity in the TG limit.
Later, it will be convenient to parameterize this as $\left<\Phi(x)\right>\equiv\Phi_0(x)+\Phi_{2\pi\rho_0}(x)\left[e^{2i\pi \rho_0x}+e^{-2i\pi\rho_0x}\right]$ and also introduce the  quantity $\Delta\equiv |g\Phi_{2\pi\rho_0}|$. The atoms will coherently scatter light into the cavity, so any nonzero density of atoms implies $\left<\Phi\right>\neq 0$. We will find that $\Phi_{2\pi\rho_0}(x)$ becomes nonzero at the Peierls transition,  leading to the dynamical generation of a mass gap. 

Adiabatic elimination of photons from our model at the mean-field level leads to completely conservative dynamics---this is a generic feature of a Rabi-like matter-light coupling~\cite{DamanetDaleyKeeling}.  The resulting conservative dynamics is determined entirely by an effective Hamiltonian, $H^{\text{eff}}$, and so the
steady-state condition becomes equivalent to  minimizing this with respect to the mean field $\Delta$.  The atomic and atom-cavity coupling parts of $H^{\text{eff}}$ come from substituting  Eq.~\eqref{Phi2} into Eq.~\eqref{H} (considering a single tube), while the cavity part can be written as $H^{\text{eff}}_{\text{cav}}= \frac{g^2}{2 \pi\hbar v_F\eta}\int \mathrm{d}x\,\left<\Phi(x)\right>^2$. We consider only constant values of $\Phi_0$ and $\Phi_{2\pi\rho_0}$, in which case $\Phi_0$  can  be absorbed into a redefinition of the chemical potential, $\mu$.  Henceforth we shall deal only with $\Phi_{2\pi\rho_0}$.  

\textit{Low energy \& Bosonization} ---
The atomic system can be described using bosonization,
which provides us with a description of our system in terms of two new bosonic fields, $\phi(x)$  and its canonical conjugate $\partial_x\theta(x)$~\cite{Coleman, Haldane, Giamarchi, GogolinNerseyanTsvelik}. The former is related to the atomic density via  $\rho(x)=\left[\rho_0-\frac{1}{\pi}\partial_x\phi(x)\right]\sum_{n=-\infty}^\infty e^{2in[\pi\rho_0-\phi(x)]}$,    while the latter is related to the current in the system~\cite{Haldane}. 
In terms of these, the steady-state condition Eq.~\eqref{Phi2} reduces to 
$g\Phi_{2\pi\rho_0}=2\pi\hbar v_F \eta\rho_0\left<\cos[2\phi(x)]\right>$.
As such, the effective mean-field Hamiltonian discussed above can be written in terms of the bosonized fields as:
\begin{multline}\label{Hbos}
H^{\text{eff}}= H^{\text{eff}}_{\text{cav}} +  \frac{\hbar v_F}{2\pi}\int\mathrm{d}x\Big\{ \frac{1}{K^2}[\partial_x\phi(x)]^2+[\partial_x\theta(x)]^2\Big\}\\
\pm 2\Delta\int\mathrm{d}x\rho_0\cos{[2\phi(x)]}. 
\end{multline}
The first line describes the  atomic and cavity systems, using the standard result for bosonization of the atoms. The atomic interactions are encoded via the parameter $K$ which depends on $U$ in a complicated fashion~\cite{Cazalilla}. For large repulsive interactions, this relationship can be approximated by $K\approx 1+ 4/ \gamma$, with the TG limit achieved at $K=1$, while $K=\infty$ corresponds to free bosons~\cite{Cazalilla}.  
The second line describes the relevant part of the matter-light coupling and will  generate a gap. We have allowed for the possibility that the  matter-light coupling and photon field might carry opposite signs, and we keep only terms which are most relevant in a renormalization group sense, which restricts our analysis to values $1/2<K<2$. 
  
While in the present work our primary system of interest is bosons with short-range, repulsive interactions,  bosonization allows one to also describe the low-energy physics of  bosons with long-range interactions or interacting fermions~\cite{Giamarchi, GogolinNerseyanTsvelik, CazalillaRMP}. Such systems are described by a Luttinger parameter  $0<K< 1$, and so in the following, we allow for arbitrary values of $1/2<K<2$. Results  for $K>1$ are applicable to bosons with short-ranged interactions or fermions with attractive interactions, whereas  $K<1$ corresponds to 1D repulsive fermions or bosonic systems with long-range interactions, the latter of which has been realized in Refs.~\cite{Tang:2018dq,Kao:2020vx}.


\textit{Tonks-Girardeau limit} ---
The atoms behave as free fermions in the TG limit~\cite{Tonks,Girardeau}. This is evident in our low-energy description at $K=1$ where it is possible to express the bosonic operators as a pair of chiral fermions.
In terms of these, the low-energy Hamiltonian Eq.~\eqref{Hbos} becomes the 1D Dirac Hamiltonian with a mass $\pm\Delta$~\cite{LutherEmery,Supplement} that is subject to the steady state condition.
This system is the same as the SSH model~\cite{SuShreiferHeeger} whose  solution is well known. Carrying it over to the present case, we find that  $\Delta=2 E_Fe^{-1/\eta} $, where $E_F=\pi\hbar v_F\rho_0$ is the Fermi energy. The self-consistent photon field is therefore given by
\begin{equation}\label{Phi_noint}
\left<\Phi(x)\right>=\pm\frac{4E_F}{g}e^{-1/\eta}\cos{(2k_Fx)}.
\end{equation}
The atomic system is insulating with a mass gap of $2\Delta$. The applicability of the low-energy description relies on $E_F$  being the largest scale in the system. In particular, we require that $\Delta<E_F$,  which in turn requires $\eta<1$.

\textit{Finite interaction} ---
The system is quite different for $K\neq 1$. It is strongly correlated and interacting, but can no longer be mapped to the SSH model. Nevertheless, an exact solution for the steady state can be found, although the cases of positive and negative $g\Phi_{2\pi\rho_0}$ need to be treated separately; we will find below that these correspond to $K>1$ and $K<1$, respectively. For $g\Phi_{2\pi\rho_0}<0$, the atomic part of the Hamiltonian is that of the Sine-Gordon or massive Thirring model with a positive mass term~\cite{Coleman, Giamarchi, GogolinNerseyanTsvelik}. This is an exactly solvable field theory and many of its properties are well known~\cite{BergknoffThacker, ZamZam}.  In particular, the mass gap of the model becomes renormalized due to the interactions, $\Delta^+_\text{R}=\xi_+ E_F\left[\Delta/E_F\right]^{1/(2-K)}$,
with $\xi_+$  a $K$-dependent constant provided in Ref.~\cite{Supplement}.
Using this, we derive the following steady-state condition from which to determine $\Phi_{2\pi\rho_0}$,
\begin{equation}\label{SC}
\frac{\Delta}{E_F}=-\eta\frac{\pi\xi^2_+}{2}\text{cot}{\left(\frac{\pi}{2-K}\right)}
\left[\frac{\Delta}{E_F}\right]^{\frac{K}{2-K}}. 
\end{equation}
This has a solution only for $K<1$. Rearranging,  we find that the self-consistent photon field and mass gap are
\begin{eqnarray}\label{Photonp}
\left<\Phi(x)\right>&=&-\frac{2E_F}{g}\zeta_+\eta^{\frac{2-K}{2-2K}}\cos{(2\pi\rho_0 x)}
,\\\label{DeltaRp}
\Delta^+_\text{R}&=&\xi_+ E_F\left[\zeta_+ \eta\right]^{\frac{1}{2-2K}},
\end{eqnarray}
where $\zeta_+$ is given in Ref.~\cite{Supplement}. The amplitude of the photon field now has a power law rather than the exponential dependence on $\eta$ in the  TG limit. Furthermore, the exponent differs from that appearing in the mass gap.
The TG limit $K\to1$ cannot be recovered from the above expression and must be treated separately as in the previous section. This highlights the strong correlations in the interacting system.  The photon field oscillates at wavevector $2\pi\rho_0$, which is $2K$ times $k_F$. 

The model maps to the massive Thirring model for $g\Phi_{2\pi\rho_0}>0$, but with a negative mass parameter. As explained in Ref.~\cite{Supplement}, this change in sign of the mass term results in the spectrum of the model being inverted; i.e., the ground state becomes the highest excited state~\cite{Supplement}. Spectral inversion also occurs when changing the sign of the interactions---i.e., taking $K\to1/K$.  Combining these, we find that for $g\Phi_{2\pi\rho_0}>0$  and $K>1$, the renormalized mass gap and self-consistent photon field are
\begin{align}\label{Photonm}
\left<\Phi(x)\right>&=\frac{2E_F}{g}\zeta_-\eta^{\frac{2K-1}{2K-2}}\cos{(2\pi\rho_0 x)},\\\label{DeltaRm}
\Delta^-_\text{R}&=E_F\xi_-\left[\zeta_- \eta\right]^{\frac{K}{2K-2}}
\end{align}
where $\xi_-, \zeta_-$ are related to $\xi_+, \zeta_+$ by $K\to1/K$.

\begin{figure}[t!]
	\includegraphics[width=\linewidth]{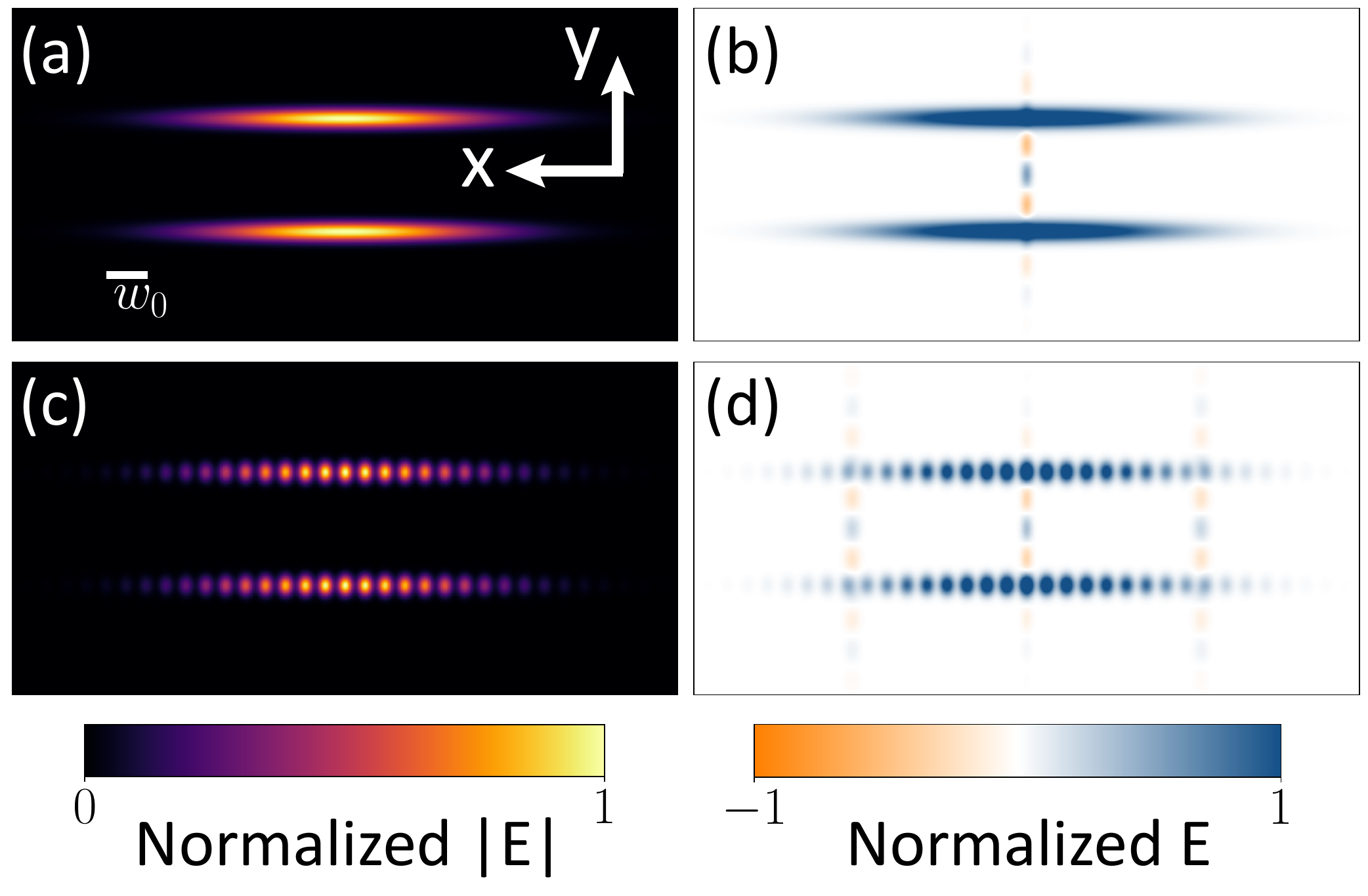}
	\caption{Left: simulated intracavity field amplitude at the atoms (a) below  and (c) far above the transition, with a density wave of wavevector $k = \pi/(0.6 w_0)$ in the TG limit. Right: cavity emission (b) below  and (d) above the transition. Emission from a confocal cavity contains both an image of the atomic density (horizontal blue pattern) and its Fourier transform (vertical blue and orange patterns).  These  can be measured by holographic reconstruction of a spatial heterodyne image~\cite{GuoKroeze,Kroeze:2018fp}. The scale bar in (a) shows the Gaussian waist $w_0$ of the cavity.} \label{Light_profile}
\end{figure}


\textit{Experimental signatures} ---
One may image the atomic density profile by using the spatial resolving power of the degenerate cavity. This provides  a  direct signature of the Peierls instability as a density wave. Figure~\ref{Light_profile} shows this, calculated using results from Ref.~\cite{GuoVaidya}. The spatial modulation of the light amplitude from the atomic image is a signature of the Peierls transition; a mirror image appears at $-y_t$. The geometry of the confocal cavity conveniently provides both the emission of  the atomic density image (long thin tubes) and its Fourier transform (vertical stripes)~\cite{GuoVaidya}. Thus, a density modulation of wavevector $2 k$ results in Bragg peaks manifest as vertical stripes.  These are positioned at $x = \pm k w^2_0$. (Each stripe is the Fourier transform of the $\hat{y}$-displaced atom image.)
 
The power-law scaling between the oscillation  amplitude of the detected light  and the parameter $\eta$ depends critically on the atom interactions; see  Eq.~\eqref{Photonp}. The Luttinger parameter $K$ can thus be experimentally measured through the dependence of this exponent on  $\eta$.  This can be tuned through its dependence on the pump intensity  $\eta\propto g^2 \propto \Omega^2$ or by the cavity-pump detuning $\omega$.
 
Probing the system by stimulating a particular photon mode realizes cavity-enhanced Bragg spectroscopy~\cite{Mottl}. In a degenerate cavity, the probe field profile can be tailored with holographic beam shaping to have a particular  wavevector. Thus, dynamic susceptibility can be measured as a function of $k$ and excitation energy by also tuning the detuning between the probe and pump. The response of the system manifests as an  increase in photon population, allowing the gap $\Delta_\text{R}^\pm$ to be measured.

In conclusion, we have shown that multimode confocal cavities can be used to realize the Peierls transition for both Bose and Fermi gases. Away from the simple limits of noninteracting fermions or TG bosons, the scaling of the detected light field with pump strength can be used to measure the Luttinger parameter. Looking beyond Peierls transitions, the compliant, phonon-supporting optical lattices inherent in multimode cavity QED make accessible a wider variety of many-body physics explorable in the context of quantum simulation.
 

Data sharing is not applicable to this article as no datasets
were generated or analyzed during the current study.
\acknowledgements  
This work was supported by US-ARO Contract
No. W911NF1310172 and Simons Foundation (V. G.),
NSF DMR-1613029 (C. R. and V. G.), and US-ARO
Contract No. W911NF1910262 (B. L.). Y. G. acknowledges funding from the Stanford Q-FARM Graduate
Student Fellowship.


\providecommand{\noopsort}[1]{}\providecommand{\singleletter}[1]{#1}%

\beginsupplement

\section{Origin of the effective Hamiltonian}

In this section, we discuss the origin of the matter-light interaction in Eq.~\eqref{H}, starting from a more general description of coupling between atoms and photons in a multimode cavity.  In general, the interaction between light and matter can be written as $H_{\text{atom,cavity}} = \int d^3\vec{r} \Psi^{3D,\dagger}(\vec{r}) \Psi^{3D}(\vec{r}) |\beta(\vec{r})|^2/\Delta_a$, where $\Delta_a$ is the atom-pump detuning, and we have written a rescaled total light field $\beta$, including the bare matter-light coupling constant.  This coupling corresponds to the standard light shift of atoms by an optical field~\cite{RitschDomokosBrenneckeEsslinger,GuoVaidya}. The total light field has two parts; that from the transverse pump along $\hat{y}$, and that from the cavity along $\hat{z}$:
\begin{equation}
\beta(\vec{r}) = \Omega \cos(k_ry) + g_0\sum_{\alpha,\nu} \hat{a}_{\alpha,\nu}
\Xi_\nu(x,y)\cos{\left[k_r\left(z+\frac{x^2+y^2}{R(z)}\right)-\theta_{\alpha,\nu}(z)\right]},
\label{totalLightField}
\end{equation}
where the Gouy phase is given by
$\theta_{\alpha,\nu}(z) = \psi(z) + n_\nu [\pi/4 + \psi(z)] - \xi_{\alpha,\nu}$, with $\psi(z) = \text{atan}\left({z}/{z_R}\right)$,
the radius of curvature $R(z) = z + z_R^2/z$, and $z_R$ the Rayleigh range.  This is  $z_R=L/2$ for a confocal cavity.  The transverse mode functions  $\Xi_\nu(x)$  are Gauss-Hermite functions of order $(l_\nu,m_\nu)$ in the $x$ and $y$ directions, respectively.

As discussed in the main text, we consider atoms trapped in tubes along  $\hat{x}$, so that we can write $\Psi^{(3D)}(\vec{r}) = \sum_t \Psi_t(x) \psi_0(y-y_t,z-z_t)$, with $\psi_0(y,z)$ describing the trapped atomic Gaussian profile of a single tube in the $yz$ plane and $y_t,z_t$ indicating the center position of each tube. We assume the trapping profile is narrow compared to the wavelength of the pump and cavity light, and that, as noted in the main text, the tubes trap atoms at the maxima of both fields, where $k_r \lambda_T/2\pi$ is an integer. 
The light shift $\propto |\beta(\vec{r})|^2$  gives, in general,  three types of terms: pump-only, cavity-only, and cross pump-cavity terms.  The pump-only terms induce a constant energy shift and so may be ignored.  We will consider the case where the bare coupling $g_0$ is much smaller than $\Omega$, so that the cavity-only term is negligible.  We can therefore focus on the cross pump-cavity term.  Restricting to points near the cavity center, we find this has the form:
\begin{equation}
    H_{\text{atom,cavity}} = \frac{\Omega g_0 }{\Delta_a}
    \int \mathrm{d}x \sum_t \Psi_t^\dagger(x) \Psi_t^{}(x)
    \sum_{\alpha,\nu} 
    (\hat{a}_{\alpha,\nu} + \hat{a}^\dagger_{\alpha,\nu})
    \tilde{\Xi}_\nu(x,y_t)
    \cos\left[-\theta_{\alpha,\nu}(0)\right].
\end{equation}
This can be rewritten directly in the form given by Eq.~\eqref{H}.
In writing this we have introduced the function:
\begin{equation}
    \tilde{\Xi}_\nu(x,y_t) = \int dy \Xi_\nu(x,y) |\psi_0(y-y_t)|^2,
\end{equation}
convolving the transverse mode functions with the trapped wavefunction.  This is important to regularise the singularity of the cavity-mediated interaction at equal positions.  Specifically, when considering the sum over modes appearing in the main text we find:
\begin{align}
\sum_{\nu}\tilde{\Xi}_\nu(x,y_t)\tilde{\Xi}_\nu(x^\prime,y_{t^\prime})&=  
\int dy |\psi_0(y-y_t)|^2 \int dy^\prime |\psi_0(y^\prime-y_{t^\prime})|^2
\sum_{\nu}{\Xi}_\nu(x,y){\Xi}_\nu(x^\prime,y^\prime)
\\&=
\int dy |\psi_0(y-y_t)|^2 \int dy^\prime |\psi_0(y^\prime-y_{t^\prime})|^2
\frac{w_0^2}{2} \left[ \delta(x-x^\prime) \delta(y-y^\prime) 
+ \delta(x+x^\prime) \delta(y+y^\prime)  \right]
\end{align}
where $w_0$ is the beam waist.  If we assume all tubes are in the upper half plane, then we need only consider the term involving $\delta(y-y^\prime)$.  Assuming a Gaussian profile $|\psi_0(y)|^2 = \exp(-y^2/2\sigma_T^2)/\sqrt{2\pi \sigma_T^2}$ with tube width $\sigma_T$ one finds the above expression reduces to:
\begin{math}
\frac{w_0^2}{2}\delta(x-x^\prime) {\delta^Y_{t,t^\prime} }/{(\sqrt{4\pi} \sigma_T)}
\end{math}
where $\delta^Y_{t,t^\prime}$ restricts to tubes at the same position $y_t$.  In the main text, for simplicity, we assume only a single tube position $y_t=y_0$, simplifying this.  For this reason, in the main text we wrote $\tilde{\Xi}_\nu(x)$ suppressing the dependence on the (constant) coordinate $y_0$.

\section{Ground-state energy of the model}
We now derive the ground-state energy density for the Sine-Gordon model with both positive and negative mass parameters. To achieve this, we work in the fermionic representation of the Sine-Gordon model, also known as the massive Thirring model~\cite{Coleman, Giamarchi, GogolinNerseyanTsvelik}. Using the right and left moving fermion operators $\psi^\dag_\pm(x)=\sqrt{\rho_0} e^{\mp i \phi(x)-i\theta(x)}$, the Hamiltonians are given by 
\begin{eqnarray}
H_\text{mtm}^\pm=\int d x\,\left\{ i\hbar v_F\left[\psi_-^\dag \partial_x\psi_--\psi_+^\dag \partial_x\psi_+\right]\pm \Delta \left[\psi^\dag_+\psi_-+\psi^\dag_-\psi_+\right]+ 4U_f\psi^\dag_+\psi^\dag_-\psi_-\psi_+ \right\},
\end{eqnarray}
where the mass parameter is $\Delta=|g\Phi_{2\pi\rho_0}|$ and $U_f$ is the fermionic  interaction strength, which is related to the Luttinger parameter $K$ in a way specified below. Therefore, $H^+_\text{mtm}$ corresponds to $g\Phi_{2\pi\rho_0}< 0$ and  $H^-_\text{mtm}$ corresponds to $g\Phi_{2\pi\rho_0}> 0$. For $U_f=0$, we have the low-energy description of the TG gas which is simply the Dirac Hamiltonain with mass $\pm\Delta$.
 
The Hamiltonian with either positive or negative mass may be solved exactly via Bethe ansatz and the many-body eigenstates determined explicitly~\cite{BergknoffThacker}. We introduce the operators $\Lambda^\dag(\theta, x)=e^{\theta/2}\psi^\dag_+(x)+e^{-\theta/2}\psi^\dag_-(x)$, which in the TG limit are merely the Bogoliubov quasiparticle creation operators for a particle with rapidity $\theta$, or equivalently, momentum $\hbar k=\Delta\sinh{(\theta)}/v_F$. In terms of these, the $N$-body wavefunctions of both  $H^+_\text{mtm}$ and $H^-_\text{mtm}$ may be written as 
\begin{eqnarray}\label{SGstate}
\int d^Nx\,\prod_{i<j}e^{i\chi(\theta_i-\theta_j)\text{sgn}(x_i-x_j)/2} \prod_{j=1}^Ne^{i\Delta\sinh{(\theta_j)x_j/\hbar v_F}}\Lambda^\dag(\theta_j,x_j)\ket{0},
\end{eqnarray}
where $\ket{0}$ is the vacuum state containing no particles and $\chi(\theta_i-\theta_j)$ is the two particle phase shift~\cite{BergknoffThacker}
\begin{eqnarray}
e^{i\chi(\theta_i-\theta_j)}=\frac{\sinh{[(\theta_i-\theta_j)/2-i\gamma]}}{\sinh{[(\theta_i-\theta_j)/2+i\gamma]}},\\
\gamma=\pi/2+\arctan{(U_f)}.
\end{eqnarray}
The two models share a set of common eigenstates, however the change in sign of the mass results in a change in the eigenvalues of these states that are $E=\pm\sum_{j=1}^N\Delta \cosh{(\theta_j)}$, with the plus sign for $H^+_\text{mtm}$ and the minus sign for $H^-_\text{mtm}$. 
The ground state consists of all the negative energy particles being filled from some cutoff up to zero energy. Therefore, the ground state of $H^+_\text{mtm}$ consists of particles whose rapidities have an imaginary part, $\theta_j=\theta^+_{j}+i\pi$, while for  $H^-_\text{mtm}$ the rapidities are purely real, $\theta_j=\theta^-_j$. In effect, the two models are related by inverting the spectrum; i.e., the ground state of one model is the highest excited state  of the other.

The rapidity parameters $\theta^\pm_j$ are not free, but instead are coupled together via the Bethe ansatz equations. These are given by~\cite{BergknoffThacker}
\begin{eqnarray}\label{Bae}
e^{\mp i\Delta\sinh{(\theta_j^\pm)L/\hbar v_F}}=\prod_{k\neq j}
\frac{\sinh{[(\theta^\pm_j-\theta^\pm_k)/2-i\gamma]}}{\sinh{[(\theta^\pm_j-\theta^\pm_k)/2+i\gamma]}}
\end{eqnarray}
and are derived by imposing periodic boundary condition on the wavefunction in Eq.~\eqref{SGstate}. We may bring the Bethe equations above into a common form by introducing $\gamma^\pm$ with $\gamma^+=\gamma$ and $\gamma^-=\pi-\gamma^+$, which in terms of the interaction strength are $\gamma^\pm=\pi/2\pm\arctan{(U_f/v_F)}$. This shows that the ground state of $H^\pm_\text{mtm}$ is described by the Bethe equations 
 \begin{eqnarray}\label{Bae2}
e^{- i\Delta\sinh{(\theta_j^\pm)L/\hbar v_F}}=\prod_{k\neq j}
\frac{\sinh{[(\theta^\pm_j-\theta^\pm_k)/2-i\gamma^\pm]}}{\sinh{[(\theta^\pm_j-\theta^\pm_k)/2+i\gamma^\pm]}}
\end{eqnarray}
and has energy $E^\pm_\text{gs}=-\sum_{j=1}^N\Delta\cosh{(\theta^\pm_j)}$. 
 
In the thermodynamic limit $N,L\to \infty$, the rapidities can be described by a distribution denoted by $\rho_\pm(\theta)$ such that the sum over rapidities is replaced by $\sum_{j=1}^N \to L\int d\theta\,\rho_\pm(\theta)$.  This can  subsequently be used to obtain the ground-state energy density $\varepsilon^\pm_\text{gs}=-\int d \theta\, \rho_\pm(\theta) \Delta\cosh{(\theta)}$. Taking the logarithm of Eq.~\eqref{Bae} and the thermodynamic limit using standard Bethe ansatz techniques (see, e.g.,~\cite{Takahashi, Thacker}), we arrive at the integral equation  for the rapidity distribution:
\begin{eqnarray}
\frac{\Delta}{2\pi \hbar v_F}\cosh{\theta}=\rho_\pm(\theta)+\int d\nu\, f_\pm(\theta-\nu)\,\rho_\pm(\nu),\\
f_\pm(x)=\frac{1}{2\pi}\frac{\sin{(2\gamma^\pm)}}{\cosh{(x)}-\cos{(2\gamma^\pm)}}.
\end{eqnarray}
The above integrals need to be regulated in some fashion to solve the integral equation. In the TG case, which is equivalent to the  SSH model, a momentum cutoff of $\pi\rho_0$ is imposed, and so we shall also employ the same strategy in the interacting case. We introduce a rapidity cutoff, $\mathcal{K}$, which is determined   via $\rho_0=\int_{-\mathcal{K}}^\mathcal{K} d \theta\, \rho_\pm(\theta)$.
Following the procedure in~\cite{Thacker}, we find that
\begin{eqnarray}
\rho_\pm(\theta)=\frac{\Delta_\text{R}^\pm}{2\pi \hbar v_F }\cosh{\left[\frac{\pi}{2\gamma^\pm}\theta\right]},
\end{eqnarray}
where $\Delta^\pm_\text{R}$ is the renormalized mass gap of the interacting model. It is related to $\Delta$ via 
\begin{eqnarray}
\frac{\Delta^\pm_\text{R}}{\pi\hbar v_F \rho_0 }&=&\xi_\pm \left[\frac{\Delta}{\pi\hbar v_F \rho_0 }\right]^{\pi/2\gamma^\pm},\\ \xi_\pm&=&\frac{\pi}{\gamma^\pm}\left[\frac{\tan{\left(\frac{\pi^2}{2\gamma^\pm}\right)}}{\pi^2/\gamma^\pm-2\pi}\right]^{\pi/2\gamma^\pm}.
\end{eqnarray}
From this, we can determine the ground-state energy density in the thermodynamic limit to be 
\begin{eqnarray}
\varepsilon^\pm_{\text{gs}}=\frac{\hbar v_F}{2}(\pi\rho_0)^2\left[\frac{2\gamma^\pm-\pi}{\pi^2/\gamma^\pm+2\pi}\right]\text{cot}{\left(\frac{\pi^2}{2\gamma^\pm}\right)}+\frac{\gamma^\pm [\Delta^{\pm}_\text{R}]^2}{2\pi \hbar v_F}\text{cot}{\left(\frac{\pi^2}{2\gamma^\pm}\right)}.
\end{eqnarray}
Note that this expression is negative only for $\gamma^\pm>\pi/3$ and so is actually only the ground state of the model within the regime $\gamma^\pm>\pi/3$, excluding the point $\gamma^\pm=\pi/2$.  That point corresponds to the TG case and needs to be separately considered. 

In order to relate these expressions back to the bosonized version of the model, Eq.~\eqref{Hbos}, we need to express $\gamma$ in terms the Luttinger parameter $K$. For the positive mass model the relation is known to be $2\gamma^+/\pi=2-K$~\cite{BergknoffThacker, ZamZam}. To find a similar relationship for the negative mass model, we use the fact that the two are related by $U_f\to -U_f$, which within bosonization is equivalent to $K\to 1/K$~\cite{Giamarchi, GogolinNerseyanTsvelik}, and therefore $2\gamma^-/\pi=2-1/K$. 

The basic excitations of $H^\pm_\text{mtm}$ consist of adding holes or particles on top of the ground states described above. The addition of a hole to the ground state distribution at $\theta^h$ leads to a shift $\rho_\pm(\theta)\to \rho_\pm(\theta)+\delta \rho_\pm(\theta)$ due to the interactions in the model. The shift is a solution to the integral equation,
\begin{eqnarray}
-\delta(\theta-\theta^h)=\delta\rho_\pm(\theta)+\int d\nu\,f_\pm(\theta-\nu)\rho_\pm(\nu),
\end{eqnarray}
which can be solved via Fourier transform. The change in energy due to the presence of the hole provides us with the single-particle excitation spectrum, i.e., the energy of the hole $\varepsilon^h(\theta^h) $:
\begin{eqnarray}
\varepsilon(\theta^h)&=&-\Delta\int d\theta \,\cosh{(\theta)}\delta\rho_\pm(\theta)=2\pi \rho_\pm(\theta^h)\\
&=&\Delta_R^\pm\cosh{\left[\frac{\pi}{2\gamma^\pm}\theta\right]}.
\end{eqnarray}
We can translate this into the more familiar language of particle momenta using $\Delta\sinh{(\theta)}=\hbar v_F k$:
\begin{eqnarray}\label{MTMholeenergy}
\epsilon^h(k)=\Delta^\pm_R\cosh{\left[\frac{\pi}{2\gamma^\pm}\text{arcsinh}{\left(\frac{\hbar v_F k}{\Delta}\right)}\right]}.
\end{eqnarray}
In the TG limit, this reproduces $\epsilon=\sqrt{(\hbar v_F k)^2+\Delta^2}$ and shows that the gap is given by $\Delta_R^\pm$. 

The excitations of the full system, both atom and cavity, consist of an atomic excitation with momentum $\hbar k$ and an associated shift of the photon field $\Delta_k'=\Delta+\delta\Delta_k$ such that 
\begin{equation}
\Delta'_k=-\pi\hbar v_F\eta \d{}{\Delta'}\left[\varepsilon_{\text{gs}}(\Delta_k')+\epsilon_k(\Delta_k')/L\right],
\end{equation}
where $\epsilon_k(\Delta_k')$ is the expression given in~\eqref{MTMholeenergy}, but evaluated at $\Delta^\prime_k$. From this, we have that to leading order, the shift in the photon field is $\delta\Delta_k=-\pi\hbar v_F\eta (\d{\epsilon_k(\Delta)}{\Delta})/L$. The total shift in energy caused by this excitation is then given by  
\begin{equation}\label{excitationenergy}
\delta E_k=L\delta\Delta_k \d{\varepsilon_{\text{gs}}(\Delta)}{\Delta}+\frac{L}{\pi\hbar v_F\eta}\delta\Delta_k\,\Delta+\epsilon_k(\Delta),
\end{equation}
where the first term in the first line comes from the shift in the ground-state energy density due to the change in $\Delta$, the second is the change in the photon energy and the last is the energy of the atomic excitation which is positive. Upon using the self-consistency condition, we find the first and second terms cancel and $\delta E_k=\epsilon_k(\Delta)$.


\section{Solution of the self-consistency equation}
As discussed in the main text, the steady-state/self-consistency condition is equivalent to minimizing the energy of the effective Hamiltonian. For both the TG and interacting cases, this becomes
\begin{eqnarray}\label{SCsupp}
\Delta=-\pi \hbar v_F\eta \d{\varepsilon^\pm_{\text{gs}}}{\Delta}.
\end{eqnarray}
Using the expression from the last section, we have that 
\begin{eqnarray}\label{derivative}
\d{\Delta^\pm_\text{R}}{\Delta}=
\frac{\pi \xi_\pm}{2\gamma^\pm}\left[\frac{\Delta}{\pi\hbar v_F\rho_0}\right]^{\pi/2\gamma^{\pm}-1}.
\end{eqnarray}
Inserting this into Eq.~\eqref{SCsupp}, we have
\begin{eqnarray}
\frac{\Delta}{\pi\hbar v_F\rho_0}=-\eta\text{cot}{\left(\frac{\pi^2}{2\gamma^\pm}\right)}\frac{\pi \xi_\pm^2}{2}\left[\frac{\Delta}{\pi\hbar v_F\rho_0}\right]^{\pi/\gamma^{\pm}-1}. 
\end{eqnarray}
The left hand side of this equation is positive by definition whereas the right hand side is positive only for $\gamma^\pm>\pi/2$: A solution is only possible within this regime. Restricting to these cases, $K<1$ for $H^+_\text{mtm}$ or $K>1$ for $H^-_\text{mtm}$,
 and rearranging, we find
\begin{eqnarray}\label{Delta}
\frac{\Delta}{\pi\hbar v_F\rho_0}=\zeta_\pm \eta^{\gamma^\pm/(2\gamma^\pm-\pi)}, 
\end{eqnarray}
where the proportionality constant is 
\begin{eqnarray}
\zeta_\pm&=&\left[\frac{\pi\xi_\pm^2}{2}\Big|\text{cot}{\left(\frac{\pi^2}{2\gamma^\pm}\right)}\Big|\right]^{\gamma^\pm/(2\gamma^\pm-\pi)}\\
&=&\frac{\pi}{\gamma^\pm}\left[\frac{\pi}{4\gamma^\pm-2\pi}\right]^{\gamma^\pm/(2\gamma^\pm-\pi)}\sqrt{\frac{\tan{\left(\frac{\pi^2}{2\gamma^\pm}\right)}}{2\gamma^\pm-\pi}}.
\end{eqnarray}
We can   express the renormalized mass parameters in terms of this result:
\begin{eqnarray}
\frac{\Delta^\pm_\text{R}}{\pi\hbar v_F \rho_0}=\xi_\pm\zeta_\pm^{\pi/2\gamma^\pm}\eta^{\pi/(4\gamma^\pm-2\pi)}.
\end{eqnarray}

\section{Realistic Experimental Parameters}
In this section, we discuss the experimental feasibility of observing the Peierls transition in a 1D system realized with bosonic $^{87}\mathrm{Rb}$ atoms. Previous experiments have achieved the Tonks-Girardeau limit by trapping $^{87}\mathrm{Rb}$ in a 2D optical lattice with sufficient lattice depth \cite{Kinoshita}. To stay within the validity of the low-energy description and maintain sufficient photon population, we require $\eta \lessapprox 1$ in Eq.~\eqref{Phi_noint}. In the experiment, the 1D tubes are tightly confined in the transverse direction, with typical excitation energy of $\sim$100 kHz, corresponding to a harmonic oscillator length scale of $\sim$30 nm. This is far below the wavelength of the cavity light. As such, the parameter $\sigma_T$ in the expression of $\eta$ is determined by the minimum spot size supported by the multimode cavity.  In previous work, the smallest spot size measured, $\sim$1~$\mu$m, was in fact limited by the atomic cloud size.  We use this estimate as a conservative upper bound. Using cavity QED parameters from Ref.~\cite{Vaidya}, we find (with a pump and cavity  $-100$ GHz detuned from the $D_2$-line of $^{87}\mathrm{Rb}$) that $\eta \approx 0.8$ is achieved with 100 tubes, 0.5$/\mu$m atomic linear density, and a pump Rabi frequency of $\sim$50 MHz at $-20$ MHz pump--cavity detuning. These conditions are realizable with existing technology. The typical temperature of the gas is around 1--10~nK, which is smaller than the mass gap $\Delta \approx k_B \cdot$15 nK. The spontaneous emission rate at such pump power and atomic detuning is $\sim$2 Hz, leaving ample time for observing the instability.

\end{document}